\documentclass[
twocolumn,
]{ceurart}

\usepackage{graphicx}
\usepackage{subcaption}
\usepackage{multirow}
\usepackage{color,soul}
\usepackage{array, caption, tabularx,  ragged2e,  booktabs}
\usepackage{tabularx}
\gappto{\UrlBreaks}{\UrlOrds}
\usepackage{subcaption}
\begin{document} \sloppy

\copyrightclause{Copyright © 2021 for this paper by its authors. Use permitted under Creative Commons License Attribution 4.0 International (CC BY 4.0).}

\conference{Joint Proceedings of the ACM IUI 2021 Workshops, April 13-17, 2021, College Station, USA}
  
\title{3D4ALL: Toward an Inclusive Pipeline to Classify 3D Contents}

\author[1]{Nahyun Kwon}[%
orcid=0000-0002-2332-0352,
]
\ead{nahyunkwon@tamu.edu}
\ead[url]{https://nahyunkwon.github.io/}
\address[1]{HCIED Lab, Texas A\&M University}

\author[1]{Chen Liang}[%
orcid=0000-0003-1645-2397,
]
\ead{cltamu@tamu.edu}

\author[1]{Jeeeun Kim}[%
orcid=0000-0002-8915-481X,
]
\ead{jeeeun.kim@tamu.edu}
\ead[url]{http://www.jeeeunkim.com/}

\begin{abstract}
Algorithmic content moderation manages an explosive number of user-created content shared online everyday. Despite a massive number of 3D designs that are free to be downloaded, shared, and 3D printed by the users, detecting sensitivity with transparency and fairness has been controversial. Although sensitive 3D content might have a greater impact than other media due to its possible reproducibility and replicability without restriction, prevailed unawareness resulted in proliferation of sensitive 3D models online and a lack of discussion on transparent and fair 3D content moderation. As the 3D content exists as a document on the web mainly consisting of text and images, we first study the existing algorithmic efforts based on text and images and the prior endeavors to encompass transparency and fairness in moderation, which can also be useful in a 3D printing domain. At the same time, we identify 3D specific features that should be addressed to advance a 3D specialized algorithmic moderation. As a potential solution, we suggest a human-in-the-loop pipeline using augmented learning, powered by various stakeholders with different backgrounds and perspectives in understanding the content.
Our pipeline aims to minimize personal biases by enabling diverse stakeholders to be vocal in reflecting various factors to interpret the content. We add our initial proposal for redesigning metadata of open 3D repositories, to invoke users' responsible actions of being granted consent from the subject upon sharing contents for free in the public spaces.
\end{abstract}


\maketitle

\section{Introduction}

To date, many social media platforms observed an explosive number of user-created content posted everyday from Twitter to YouTube to Instagram and more. 
Following the acceleration of online contents which becomes even faster partly due to COVID-19, it has also become easier for people to access sensitive content that may not be appropriate for the general purpose. 
Owing to the scale of these content and users' abilities to share and repost them in a flash, it becomes extremely costly to detect the sensitive content solely by manual work.
Current social media platforms have adopted various (semi)automated content moderation methods including a deep learning-based classification (e.g., Microsoft Azure Content Moderator \cite{MSAzure}, DeepAI's Nudity Detection API \cite{DeeapAINudity}, Amazon Rekognition Content Moderation \cite{rekognition_moderation}). 

Meanwhile, since desktop 3D printers have been flooded into the consumer market, 3D printing specific social platforms such as Thingiverse \cite{thingiverse} have also gained popularity, contributing to the proliferation of shared 3D contents that are easily downloadable and replicable among community users.
Despite a massive number of 3D contents shared for free to date---As of 2020 2Q, there are near 1.8 million 3D models available for download, excluding empty entries due to post deletion---, there has been relatively little attention to sensitive 3D contents. 
This might result in not only a lack of a dataset to be used as a bench mark, but also 
a lack of discussion on fair rationales to be utilized in building a algorithmic 3D content moderation that integrates everyone's perspectives with a different background.
Along with significant advances in technology of machine mechanisms and materials (e.g., 3D printing in metals), the 3D printing community may present an even greater impact from the spread of content due to its limitless potential for replication and reproduction.
In view of various stakeholders who have different perspectives in consuming and interpreting contents
---from K-12 teachers who may seek 3D files online to design curricula to artists who depict their creativity in digitized 3D sculptures---, moderating 3D content with fairness becomes more challenging.
3D contents online often consist of images and text that are possibly useful to adopt existing moderation schemes including text (e.g., \cite{prabowo2009sentiment, baccianella2010sentiwordnet, saha2018hateminers, Ahluwalia2018hateSpeech}) or image based (e.g., \cite{minaee2019machine, kumar2020image, llanso2020artificial}) approaches.
However, there exist 3D printing specific features (e.g., print support to avoid overhangs, uni-colored outcome, segmented in parts, etc.) that may prevent direct adoption of those schemes, requiring further consideration about implementing advanced 3D content moderation techniques.


In this work, we first study the existing content moderation efforts that has potential to be used in 3D content moderation
and discuss shared concerns in examining transparency and fairness issues in algorithmic content moderation.
As a potential solution, we propose a semi-automated human-in-the-loop validation pipeline using augmented learning that incrementally trains the model with the input from the human workforce. 
We highlight potential biases that are likely to be propagated from different perspectives of human moderators who provide final decisions and labeling for re-training a classification model.
To mitigate those biases, we propose an image annotation interface to develop an explainable dataset and the system that reflects various stakeholders' perspectives in understanding the 3D content.
We conclude with initial recommendations for metadata design to (1) require consent and (2) inform previously unaware users of consent for publicizing the content which might invade copyright or privacy.

\section{Algorithmic Content Moderation}
Manual moderation relying on a few trusted human workforce and voluntary reports has been common solutions to review shared contents.
Unfortunately, it becomes increasingly difficult to meet the demands of growing volumes of users and user-created content \cite{crowdmoderation}. 
Algorithmic content moderation has taken an important place in popular social media platforms to prevent various sensitive content in real-time, including graphic violence, sexual abuse, harassment, and more.
As with other media posts
, 3D contents available online appear as web documents that consist of images and text.
For example, to attract audiences and help others understand the design project, creators in Thingiverse voluntarily include various information such as written descriptions of the model, tags, as well as photos of a 3D printed design; thus, 3D content can provide us an ample opportunity to employ the existing text and image based moderation schemes.


Among various text-based solutions, sentiment analysis is one traditionally popular approach that categorizes input text into either two or more categories: positive and negative, or more detailed \textit{n}-point scales (e.g., highly positive, positive, neutral, negative, highly negative) \cite{prabowo2009sentiment, baccianella2010sentiwordnet}. 
Moderators can consider categorization results in deciding whether the content is offensive or discriminatory \cite{SentimentAnalysis}. 
Various classifiers, such as Logistic Regression Model, Support Vector Machine, and random forest, are actively used in detecting misogynistic posts on Twitter (e.g., \cite{saha2018hateminers, Ahluwalia2018hateSpeech}).
Jigsaw and Google's Counter Abuse Technology suggested Perspective API \cite{perspective_api} provide a score on how \textit{toxic} (i.e., rude, disrespectful, or unreasonable) the text comment is, using a machine learning (ML) model that was trained by people's rating of internet comments.

With the rapid improvement of Computer Vision (CV) technologies with machine learning, several image datasets (e.g., NudeNet Classifier dataset\cite{nudenet_dataset}) and moderation APIs enable developers to apply these ready-to-use mechanisms to their applications. 
For example, Microsoft Azure Content Moderator \cite{MSAzure} classifies adult images into several categories, such as explicitly sexual in nature, sexually suggestive, or gory. 
DeepAI's Nudity Detection API \cite{DeeapAINudity} enables automatic detection of adult images and adult videos.
Amazon Rekognition content moderation \cite{rekognition_moderation} detects inappropriate or offensive features in images and provides detected labels and prediction probabilities.
However, many off-the-shelf services and APIs are often obscured, because it is hard for users to expect that the models are trained with fair ground-truths that can offer reliable results to various stakeholders with different cultural or social backgrounds without any biases, which we will discuss more in a detailed way in the following section.




\subsection{Challenges in Moderating 3D Content}
As we noted earlier, 3D contents appear as web documents that consist of text descriptions, auto-generated preview images, and user-uploaded images to help others comprehend the content at a glance. 
Although it is technically possible to utilize existing text and image based moderation schemes, 3D models have unique features that make it hard to directly adopt the existing CV techniques to their rendered images or photos. 
\subsubsection{3D specific features that hamper the use of existing CV techniques}
We identified four characteristics that make sensitive elements undetectable by the existing algorithms.


\begin{figure*}%
    \centering
    \subfloat[\centering Rotated model]{{\includegraphics[width=.5\columnwidth]{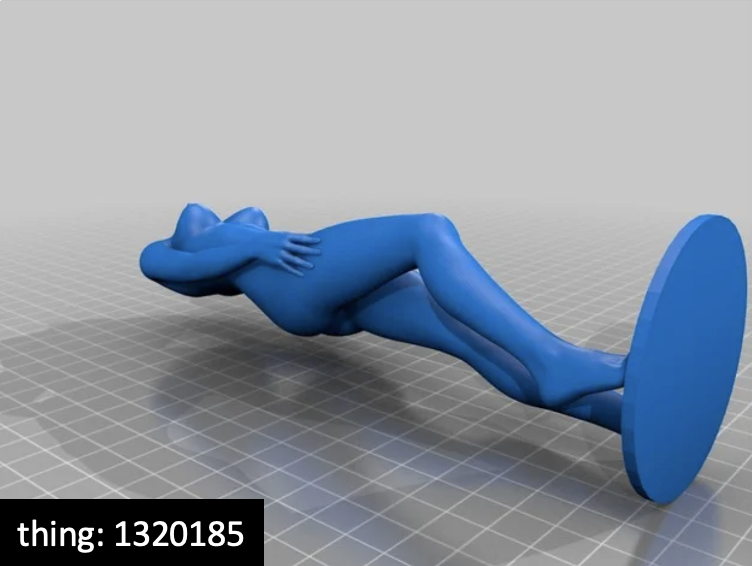} }}%
    \subfloat[\centering Support structure]{{\includegraphics[width=.505\columnwidth]{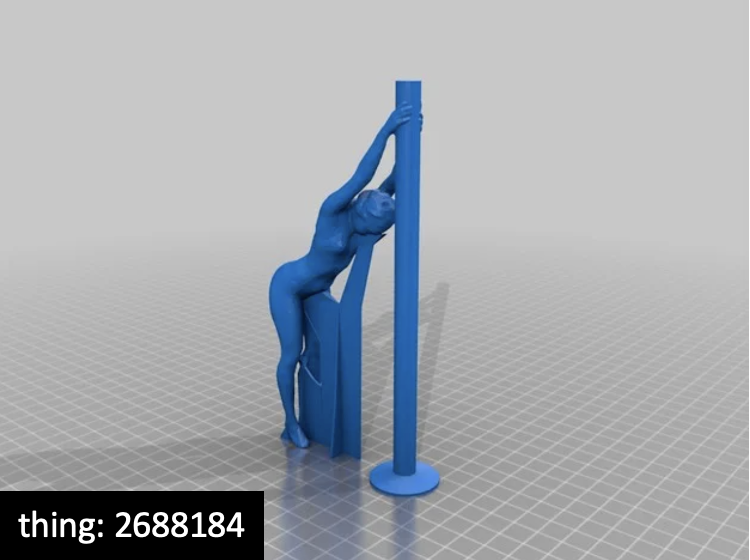} }}%
    \subfloat[\centering Texture on surface]{{\includegraphics[width=.494\columnwidth]{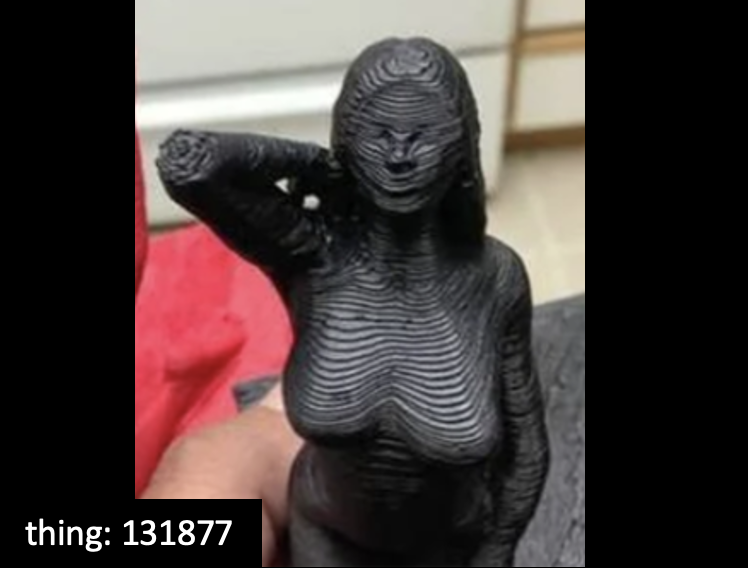} }}%
    \subfloat[\centering Divided into parts]{{\includegraphics[width=.498\columnwidth]{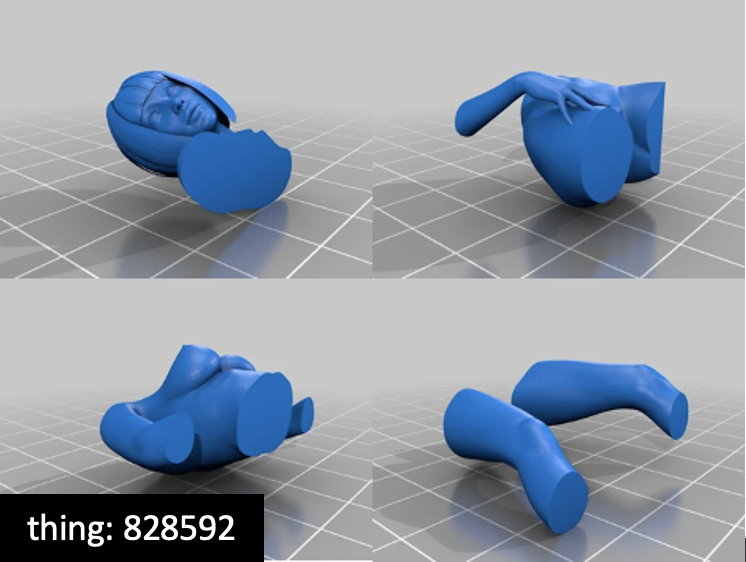} }}%
    \caption{Example images for the mainly 4 characteristics that make it hard to use the existing CV techniques; each thing is reachable using its unique ID through the url of https://thingiverse.com/thing:ID}%
    \label{fig:challenges}%
\end{figure*}

\noindent
\textbf{Challenge 1. Difficulties in Locating Features from Images of the Current Placement.}
Thingiverse automatically generates rendered images of the 3D model when a 3D file is uploaded, and this is used as a representative image if the designer does not provide any photos of real 3D prints.
In many cases, these files are placed in the best orientation that guarantees print-success in FDM (Fused Deposition Modeling) printers, aligning the design to minimize overhangs.
As the preview is taken in a fixed angle, so it might not be in a \textit{perfect} angle that shows the main part of the model thoroughly (e.g., Fig~\ref{fig:challenges}(a)). 
It hinders 
the existing image-based APIs from accurate detection of sensitivity in the preview images, because sensitive parts might not be visible.

\vspace{2.0ex}
\noindent\textbf{Challenge 2. Support Structure that Occludes the Features.}
Following the model alignment strategy of FDM printing, designers often include a custom support structure to prevent overhangs and to avoid printing failures and deterring surface textures with auto-generated supports from slicers (i.e., 3D model compiler) such as Cura \cite{cura}. 
These special structures easily occlude the design's significant features (e.g., Fig~\ref{fig:challenges}(b)). 
Since the model is partly or completely occluded, the existing CV techniques barely detect sensitivity of the design.

\vspace{2.0ex}
\noindent\textbf{Challenge 3. Texture and Colors.}
Current 3D printing technologies enable users to use various print settings and other postprocessing techniques. 
Accordingly, the printed model may present unique appearances compared to general real-world entities.
Often the model is single-colored and can have a unique texture such as linear lines on the surface (e.g., Fig~\ref{fig:challenges}(c)) due to the nature of 3D printing mechanisms of accumulating materials layer-by-layer, which might let the existing CV algorithms overlook the features.

\vspace{2.0ex}
\noindent\textbf{Challenge 4. Models Separated into Parts for Printing.}
As one common 3D printing strategy to minimize printing failures from a complex 3D designs such as a human body, many designers divide their models into several parts to ease the printing process, and let users post-assemble as shown in Fig~\ref{fig:challenges}(d). 
In this case, it is hard for the existing CV techniques to get the whole assembled model, resulting in a failure to recognize its sensitivity.


\section{Transparency and Fairness Issues in Content Moderation}

\subsection{Transparency: Black Box that Lacks Explanation}

Content moderation has long been controversial due to its non-transparent and secretive process \cite{gorwa2020algorithmic}, resulting from lacking explanations for community members about how the algorithm works. 
To meet the growing demands for transparent and accountable moderation practice as well as to elevate public trust, recently, popular social media platforms have begun to dedicate their efforts to make their moderation process more obvious and candid \cite{gorwa2020algorithmic, granados2013transparency, transparency_democracy, maccarthy2020transparency}.
As a reasonable starting point, those services provided detailed terms and policies (e.g., Facebook's Community Standards \cite{facebook_community}) describing the bounds of acceptable behaviors on the platform \cite{gorwa2020algorithmic}. 
In 2018, as a collective effort, researchers and practitioners proposed the Santa Clara Principles on Transparency and Accountability in Content Moderation (SCP) \cite{scp}. 
SCP suggests one requirement that 
social media platforms should provide detailed guidance to the members about which content and behaviors are discouraged, including examples of permissible and impermissible content, 
as well as an explanation of how automated tools are used across each category of content.
It also recommends for content moderators to give users a \textit{rationale} for content removal to assure about what happens behind the content moderation.

Making the moderation process transparent and explainable is crucial to the success of the community \cite{juneja2020through}, in order not only to maintain its current scale but also to invite new users, because it may affect users' subsequent behaviors.
For example, given no explanation about the content removal, users are less likely to upload new posts in the future or leave the community, because they may believe that their content was treated unfairly thus get frustrated owing to an absence of communication \cite{jhaver2019did}.
Reddit \cite{Reddit}, which is one of the most popular social media, has equipped volunteer-based moderation schemes resulting in the removal of almost one fifth of all posts every day \cite{jhaver2018did} due to violation of their community policy \cite{reddit_policy} (e.g., Rule 4: \textit{Do not post or encourage the posting of sexual or suggestive content involving minors.}) or individual rules of the subreddits (i.e., subcommunity of Reddit that has a specific individual topic) according to their own objectives (e.g., One of the rules in 3D printing subreddit: ``Any device/design/instructions which are \textit{intended} injure people or damage property will be removed.''). 
Users being aware of community guidelines or receiving explanations for content removal are more likely to perceive that the removal was fair \cite{jhaver2019did} and showcase more positive behaviors in the future.
As many social platforms including 3D open communities such as Thingiverse highly rely on voluntary posting of the user-created content \cite{ccomlekcci2019custodians}, the role of a transparent system in content moderation becomes more significant in maintaining the communities themselves.

Even if many existing social media platforms have their full gears to implement artificial intelligence (AI) in content moderation, it has long been in the black box 
\cite{juneja2020through}, thus not understandable for users due to the complexity of the ML model.
To address the issue of the uninterpretable model that hinders the users from understanding how it works, researchers shed lights on the blind spot by studying various techniques to make the model \textit{explainable} (e.g., \cite{letham2015interpretable, wang2017bayesian, freitas2014comprehensible}).
Explainability has been on the rise to be an effective way of enhancing transparency of ML models \cite{lepri2018fair}. 
In order to secure explainability, the system must enable stakeholders to understand the high-level concepts of the model, the reasoning used by the model, and the model's resulting behavior \cite{bhatt2020explainable}. 
For example, as shown in the Fairness, Accountability, and Transparency (FAT) model, supporting users to know which variables are important in the prediction and how they will be combined is one powerful way to enable them to understand and finally trust the decision made by the model \cite{lakkaraju2016interpretable}.

\subsection{Fairness: Implicit Bias and Inclusivity Issues}
People often overlook fairness of the moderation algorithm and tend to believe that the systems automatically make unbiased decisions \cite{garfinkel2017toward}. 
In fact, the human adjudication of user-generated content has been occurred in secret and for relatively low wages by unidentified moderators \cite{roberts2019behind}. 
In some platforms, users are even unable to know the presence of moderators or who they are \cite{roberts2016commercial}, and thus it is hard for them to know what potential bias, owing to different reasoning processes, has been injected into the moderation procedure. 
For example, there have been worldwide actions that strongly criticize the sexualization of women's bodies without inclusive inference (e.g., `My Breasts Are Not Obscene' protest by the global feminist group Femen \cite{femen} to denounce a museum's censorship of nudity.).
Similarly, Facebook's automatic turning down of postings and selfies that include women's topless photo by tagging them as \emph{Sexual/Porn} ignited `My Body is not Porn' movement \cite{MyBodyNotPorn, facebook_porn}.
The different points of view in perceiving and reasoning towards the same piece of work makes it yet hard to decide the absolute sensitivity. 
It is nearly impossible that the sole group of users represent all, therefore, it is difficult for users to expect a \textit{ground-truth} in the decision-making process, and trust the result while believing experts made the final decisions based on thoughtful consideration with an unbiased rationale. 

Subsequently, many studies (e.g., \cite{blodgett2016demographic, binns2017like}) have explored potential risks of algorithmic decision-making that are potentially biased and discriminatory to a certain group of people such as underrepresented groups of gender, race, disability.
Classifier has been one common approach in content moderation, but developing a perfectly fair set of classifiers in content moderation is complex compared to those in common recommendation or ranking systems, as classifiers tend to inevitably embed a \textit{preference} to the certain group over others to decide whether the content is offensive or not \cite{gorwa2020algorithmic}.

\subsection{Transparency \& Fairness Issue in 3D Content Moderation}
Through a text feature based classification, we identified there are three main categories of sensitive 3D content: (1) sexual/suggestive, (2) dangerous weaponry, and (3) drug/smoke. 
Due to the capability of unlimited replication and reproduction in 3D printing, unawareness of these 3D contents could be crucial.
We noticed that Thingiverse limits access to \textit{some} of sensitive things that are currently labeled as NSFW (Not Safe for Work) by replacing their thumbnail images with the black warning images.
It is a secretive process because there are no clear rationale or explanations offered to users behind this process.
Therefore, users cannot expect whether Thingiverse operates based on an unbiased and fair set of rules. 

While the steep acceleration of increments of 3D models \cite{thingiverse2018celebrates} is making automatic detection of sensitive 3D content imperative,
moderating 3D content also faces fairness issues and users are suffering from lacking explanations.
We need to take our account into various stakeholders' points of view that affect their decision on potentially sensitive 3D content, as well as further discussions to mitigate bias and discrimination of the algorithmic decision-making system.
Here we propose an explainable human-in-the-loop 3D content moderation system to enable various users who have distinct rules to participate in calibrating algorithmic decisions to decrease bias or discrimination of the algorithm itself.
Although we focus on specific issues in shared 3D content online, our proposed pipeline generally applies to advancing a semi-automatic process toward an explainable and fair content moderation for all. 

\section{Towards Explainable 3D Moderation System}

A potential solution to examine 3D contents' sensitivity with fairness is employing the human workforce with ample experiences in observing and perceiving with various perspectives.
We suggest a human-in-the-loop pipeline, based on the idea of incremental learning \cite{RN15} that the human workforce can collaborate with an intelligent system, concurrently classifying data input and annotate features with the explanation for the decision.  

\subsection{Building an Inclusive Moderation Process}
Making decisions on the sensitivity of a 3D model can be subjective due to various factors such as cultural differences, the nature of the community, and the purpose of navigating 3D models. 
To reflect different angles in discerning the nature and intention of contents, we need to deliberate various interpretations taken from various groups of people. 
For example, there are lots of 3D printable replicas of artistic statues or Greek sculptures that are reconstructed by 3D scanning of the original in the museums \cite{openculture}.
Speculative K-12 teachers designing their STEAM education curriculum using 3D models are not likely to want any NSFW designs revealed to their search results.
On the other hand, there are many activists and artists who may want to investigate the limitless potential of the technology, sharing a 3D scanned copy of the naked body of herself \cite{vice} or digitizing nude sculptures available in the museum to make the intellectual assets accessible to everyone, etc. 
The nude sculpture has been one popular form of artistic creation in history, and it is not simple to stigmatize these works as `sensitive'.
Everyone has their own right to `leave the memory of self' in a digital form.
Forcing to adapt a preset threshold of sensitivity and filter these wide array of user-created contents could unfairly treat one's creative freedom. 
As the extent that various stakeholders perceive the sensitivity could be distinct,
our objective is to design an inclusive process in accepting and adopting the sensitivity.


\subsection{Solution 1: Human-in-the-loop with Augmented Learning}
Automated content moderation could help review of a vast amount of data and provide filtered cases for humans to support a decision-making process \cite{jhaver2019did}, if we well-echo diverse perspectives in understanding contents. 
\begin{figure*}%
    \centering
    \subfloat[\centering Human-in-the-loop pipeline]{{\includegraphics[width=\columnwidth]{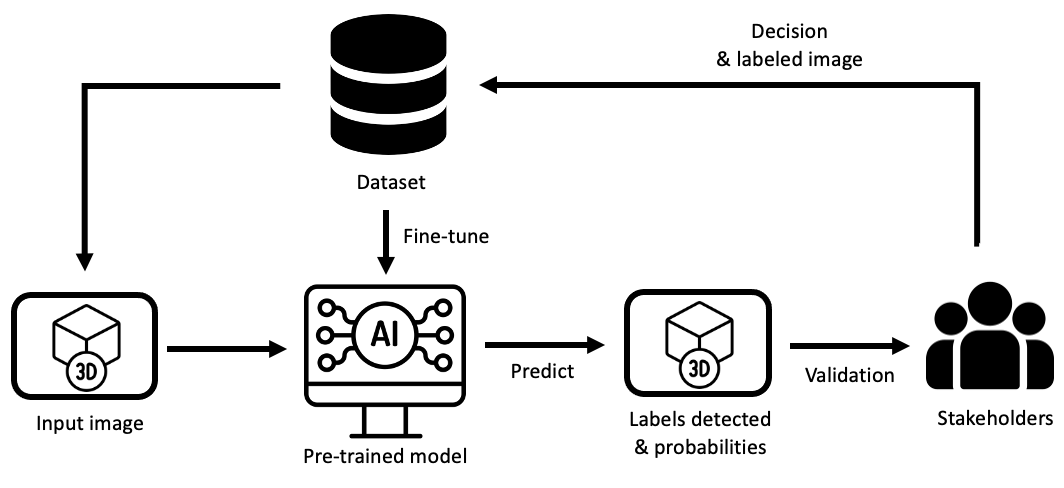} }}%
    \subfloat[\centering User interface mockup]{{\includegraphics[width=\columnwidth]{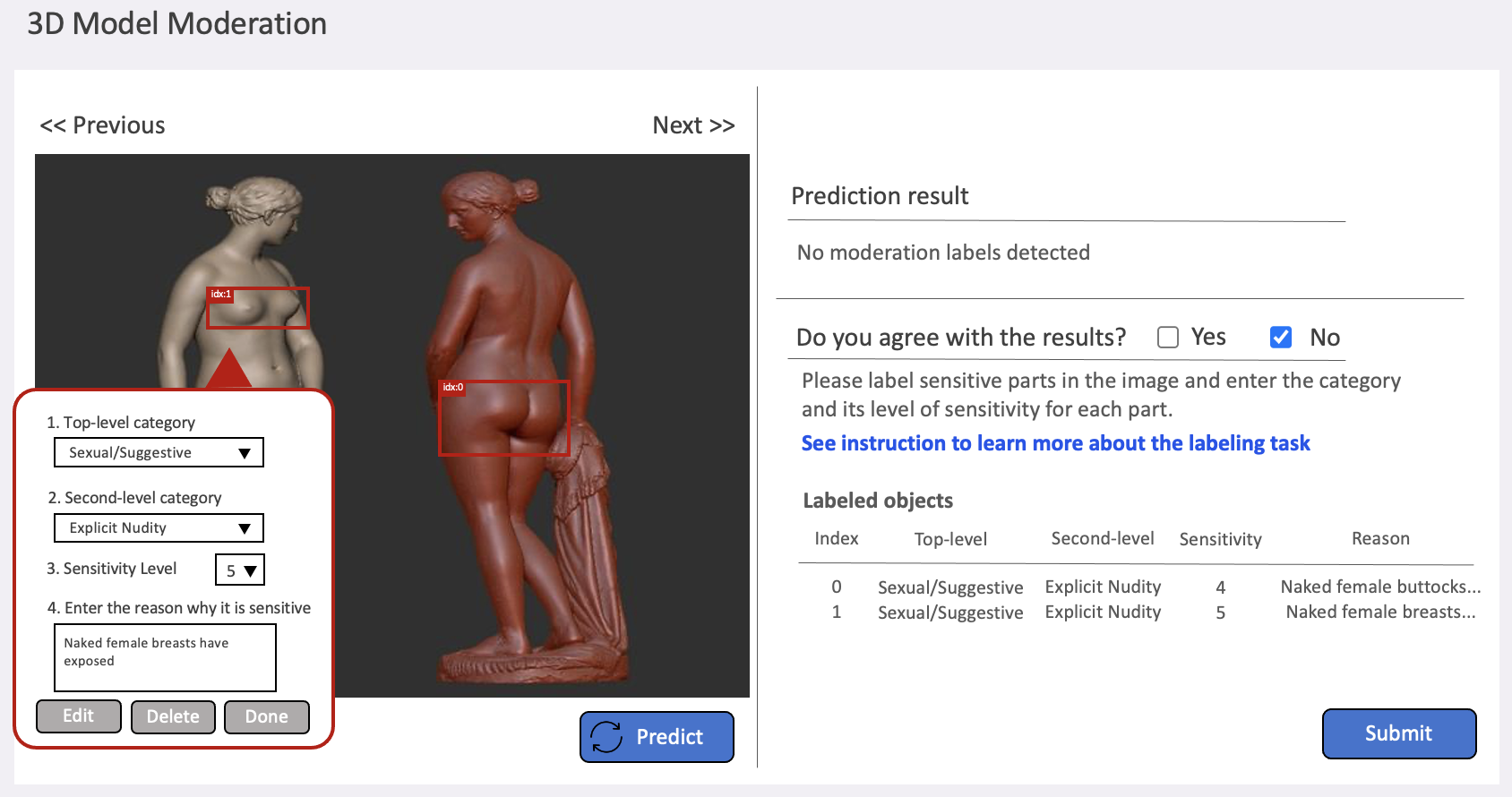} }}%
    \quad
    \caption{(a) Overview of the human-in-the-loop pipeline powered by human moderators to acknowledge various perceptions of sensitivity and (b) an user interface mockup for the moderators to validate prediction results and provide annotations regarding their rationale, thus to augment the model.}%
    \label{fig:pipeline_interface}%
\end{figure*}
In our proposal of the human-in-the-loop pipeline (Fig~\ref{fig:pipeline_interface}(a)), an input image dataset of 3D models will be used for the initial model training, then the result will be reviewed by multiple human moderators step by step.
We trained the model with 1,077 things that are already labeled as NSFW by Thingiverse and 1,077 randomly selected non-NSFW things.
All input images are simply categorized as NSFW or not, with no annotation for specific image features to provide the reasoning. 
Human moderators recruited from various groups of people now review the classification results whether they agree.
They are asked to annotate image segments using a bounding box where they referred to make the final decision with the category.
At the same time, they provide the rough level of how much the part affected the entire sensitivity and a written rationale for the decision.
These features will enhance the data quality so to be used to fine-tune the model with the weighted score, thus the model becomes able to recognize previously unknown sensitive models based on the similarity and now can \textit{explain} sensitive features.

When two different groups of people with different standards do not agree on the same model's classification results, 
the model uses their decision, annotated features, and levels of sensitivity to differentiate the extent of perceived sensitivity and reflect to the different threshold. 
For example, one moderator thinks that the model is sensitive while the other does not, the model will have a higher threshold in categorizing the content.
Different decisions on the same model finally could be brought to the table for further discussion if needed, for example, to regulate policy guidelines, or used as search criteria for other community users who have similar goals in viewing and unlocking analogous 3D contents. 
To summarize, one iteration contains the following steps: 
\begin{enumerate}
  \item The pre-trained model presents prediction results.
  \item The human moderator can enter disagreement/agreement with the results and annotate sensitive parts with a sensitivity level and a decision rationale.
  \item The annotated image is used to fine-tune the model.
  \item If the decision for the image is different from other moderators, annotations and sensitivity levels are used to set the different threshold.
\end{enumerate}

We elaborate more on feedback from the moderators by showing three possible scenarios: (1) the moderator's agreement with the prediction results, (2) sensitive parts not detected, and (3) false-classification of insensitive features sensitive. 

\noindent\textbf{Case 1. Agreement with the Prediction Result}
In case that the moderators agree with the decision, they can either finalize it or reject the classification,
by selecting provided top-level categories (e.g., sexual/suggestive, weaponry, drug/smoke) and second-level categories (e.g., under sexual/suggestive, explicit nudity, adult toys, sexual activity, etc.). 
We currently refer to a two-level hierarchical taxonomy of Amazon Rekognition to label categories of inappropriate or offensive content.

\noindent\textbf{Case 2. Sensitive Parts Ignored by the Algorithm}
Another possible case is that the specific feature in the image that the moderator perceives as sensitive is missing in the detection results. 
In this case, human moderators can label that part and provide rationales using \textit{enter the level of sensitivity} field from 1 (slightly sensitive) to 5 (highly sensitive), how each specific part affects the entire sensitivity of the model. 

\noindent\textbf{Case 3. False Negative}
It is also possible that some parts detected by the model are not sensitive for the moderator due to the higher tolerance to sensitivity.
The moderator can either submit the disagreement or provide more detailed feedback by excluding specific results. 

Different \textit{degrees} of sensitivity perception from various stakeholders can reflect distinct points of view, which may manifest fairness in algorithmic moderation through multiple iterations of this process.
In our interface for the end-users that assists searching 3D designs, we let users set their desired threshold.
For those who might find it difficult to decide a threshold that perfectly fits their need, we show several random example images that have detected sensitive labels with the corresponding threshold.
This pipeline also helps obtain the explainable moderation algorithm.
Our model can help users understand the rationales of the model by locating detected features/prediction probabilities in the image and providing written descriptions that the moderators entered for data classification.



\subsection{Solution 2: New Metadata Design to Avoid Auto-Filtering}

Another potential problem in open 3D communities is copyright or privacy-invasive contents that are immediately marked as NSFW by Thingiverse indicating they are \textit{inappropriate}.
Currently, Thingiverse lacks notification and explanation for content removal, while a majority of them might invade copyrights. 
Its obscurity results in a negative impact on the user's future behaviors.
For example, creators are frustrated at the un-notified removal of their content thus decided to quit their membership (e.g., \cite{quit_thingiverse}), which might not happen if they saw an informative alert when they post the content. 
Along with advanced 3D scanning technologies 
\cite{all3dp}, many creators are actively sharing 3D scanned models (e.g., As of December 2020, Thingiverse has 1150 things that tagged with `3D\_scan' and 308 things with the tag `3D\_scanning'). 
With arising concerns over possible privacy invasion in sensitive 3D designs, what caught our attention is 3D scanned replicas of human bodies. 
Many of them do not include an explicit description of whether the creator received the consent from the subject (e.g., \cite{scan_amber, scan_mel}).
Some designers quoted the subject's permission, 
for example, one creator describes that the subject, Nova, has agreed to share her scanned body on Thingiverse \cite{scan_nova}.
Still, this process relies on the users' voluntary action given no official guidelines, resulting in a lack of awareness that the users must be granted the consent to upload possibly privacy-invasive contents at the time of posting those content in public spaces regardless of the commercial purpose.
Without explicit consent, the content is very likely to be auto-filtered by Thingiverse, which decreases fairness by hampering artistic/creative freedom. 
To iron out a better content-sharing environment in the these open communities, redesigning of metadata must be considered and adapted by system admins that invoke responsible actions.
For example, providing a checkbox that asks \textit{``If the design is made of 3D scanned human subject, I got an agreement from the subject''} can inform previously unaware users about the need for permission to post potentially privacy-breaching contents.
Including the subject's consent can also protect creative freedom from auto-filtering, by adding that the content is not breaching copyright or privacy and can be shared in the public spaces. 
In addition, it can enable users to understand that an absence of consent could be the reason for filtering.

\section{Conclusion}
As an inclusive process to develop transparent and fair moderation procedure in 3D printing communities, our study proposes to build an explainable human-in-the-loop pipeline.
We aim to employ diverse group of human moderators to collect their rationales, which can be used to enhance the model's incremental learning.
Our objective is not to censor 3D content but to build a pleasant 3D printing community for all, by safeguarding search as well as guaranteeing creative freedom, through the pipeline and new metadata design that has potential to minimize issues related with privacy or copyright.

\bibliography{references.bib}

\end{document}